\documentstyle[prl,twocolumn,aps,epsf,rotate,bookmath]{revtex}
\date{March 6, 1997; to appear in PRL}
\begin{document}
\twocolumn[\hsize\textwidth\columnwidth\hsize\csname @twocolumnfalse\endcsname

\title{
Tunneling into Current-Carrying Surface States of High T$_c$ Superconductors}
\author{M.~Fogelstr\"om$^{a}$, D. Rainer$^{b}$ and J.~A.~Sauls$^{a}$}
\address{$^a$Department of Physics \& Astronomy, Northwestern University,
Evanston, IL 60208\\
$^b$Physikalisches Institut, Universit\"at Bayreuth, Bayreuth, Germany}
\maketitle
\begin{abstract}
Theoretical results for the $ab$-plane tunneling conductance
in the $d$-wave model for high T$_c$ superconductors are presented.
The $d$-wave model predicts surface bound states
below the maximum gap. 
A sub-dominant order parameter, stabilized by the surface,
leads to a splitting of 
the zero-bias conductance peak (ZBCP) 
in zero external field
and to spontaneous surface currents.
In a magnetic field screening currents shift
the quasiparticle bound state spectrum and
lead to a voltage splitting of the ZBCP
that is linear in $H$ at low fields,
and saturates at a pairbreaking critical field of order
$H_c^*\simeq 3\,T$.
Comparisons with recent experimental results on
Cu/YBCO junctions are presented.
\end{abstract}
\vspace*{0.2 cm}
] 


In the $d_{x^2-y^2}$ model for the cuprate superconductors surface states are
predicted which should be observable in the sub-gap conductance
for $ab$-plane tunneling.\cite{hu94,tan95,buc95}
Observations of a zero-bias peak (ZBCP) in the in-plane conductance were
reported for tunnel junctions on oriented YBCO films by
Geerk, et al.\cite{gee88}, Lesueur et al.\cite{les92},
and Covington et al.\cite{cov96}
The ZBCP splits in a magnetic field of a few Tesla, and recent 
experiments show that a splitting of a few $meV$
also appears at low temperatures
in zero-field.\cite{cov97}
The identification of the ZBCP with surface bound-states 
associated with d-wave pairing is important because the
origin of the ZBCP in the $ab$-plane tunneling conductance is the
same as that of the $\pi$ phase shift in the Josephson interference
experiments.\cite{v-har95}
The same sign change that leads to a $\pi$ phase shift
in the Josephson current-phase relation is also responsible for a ZBCP
in the $ab$-plane quasiparticle tunneling conductance for high
impedance junctions.\cite{hu94,tan95,buc95}
The ZBCP is observed with
comparable magnitude for both $(110)$ and $(100)$ orientated surfaces,\cite{cov96}
whereas theoretical calculations\cite{hu94,tan95,buc95} 
predicted no ZBCP for $(100)$ interfaces.
We show that a large ZBCP is expected in the $ab$-plane tunneling conductance
for all orientations if the interface is microscopically rough.

Several authors have suggested that the surface of
d-wave superconductors might exhibit spontaneously generated surface currents
associated with the presence of a second superconducting 
order parameter.\cite{sig95,mat95}
The surface state of any
$d_{x^2-y^2}$ superconductor will exhibit a spontaneously
broken time-reversal symmetry
phase at sufficiently low temperature (Ref.\onlinecite{buc95} and below).
The tunneling spectrum for current-carrying states of d-wave superconductors,
externally imposed and spontaneously generated, is
the main subject of this letter.

When an excitation reflects elastically off a $(110)$
surface its momentum
changes, ${\bf p}_f\rightarrow\underline{\bf p}_f$.
Incident and reflected wavepackets propagate
through different order
parameter fields, $\Delta({\bf p}_f,{\bf R})$ vs.
$\Delta(\underline{\bf p}_f,{\bf R})$, which
leads to Andreev scattering, a process of ``retro-reflection'' in which a
particle-like excitation undergoes branch conversion into a hole-like
excitation with reversed group velocity. Bound states occur at
energies for which the phases of Andreev-reflected particle- and
hole-like excitations interfere constructively. This effect is pronounced
if the scattering induces a change in sign of the order parameter along
the classical trajectory. In this case a zero-energy bound state forms.
Surface bound states of this origin were
discussed in the context of tunneling into unconventional
superconductors by Buchholtz and Zwicknagl,\cite{buc81}
and more recently for the $d_{x^2-y^2}$ pairing model
of the cuprates.\cite{hu94,tan95,buc95} 
Reversal of the velocity by Andreev reflection
is accompanied by a change in sign of the charge. Consequently
Andreev bound states can carry current. Charge conservation is
maintained by conversion of the bound state current into bulk supercurrents
far from the interface. This is the physical reason that 
$ab$-plane tunneling currents are expected to exhibit large zero-bias
conductance peaks and to be a sensitive probe of surface states
associated with the d-wave order parameter.

The physics of the surface currents and their influence 
on the subgap conductance
requires a microscopic theory of the surface 
excitations, the surface order parameter and a transport
theory for the coupled surface and bulk excitations
and the inhomogeneous pair condensate.
We consider low-transmission tunnel junctions between a normal electrode
and a cuprate superconductor oriented for tunneling into the $ab$-plane.
At $T=0$ the tunneling conductance measures the excitation spectrum of
the superconductor at the surface,\cite{duk69}
$dI/dV=\frac{1}{R_N}\int_{\vp_f\cdot\hat{\vn}>0}\,d^2p_f\,{\cal D}(\vp_f)\,N(\vp_f,\vR_{s}; eV)$,
where $R_N$ is the interface resistance in the normal-state, and
$N(\vp_f,\vR_{s};\epsilon)$ is the superconducting density of states
at the interface ($\vR_s$) for trajectories defined by the
Fermi surface position, $\vp_f$.
At finite temperatures
the equation for $dI/dV$ must be modified to reflect the thermal occupation
of the states contributing to the tunneling current.
The tunneling conductance is obtained by folding the
angle-resolved density of states at the surface
with the barrier transmission
probability, ${\cal D}(\vp_f)$, for tunneling
characteristics labeled by $\vp_f$.
We model the tunneling barrier with a uniform probability distribution,
${\cal D}=1/2\phi_c$, within an acceptance cone of angle
$2\phi_c$ about the interface normal, 
and zero outside the cone. A small value of $\phi_c$ represents to
a thick tunneling barrier, while a large value of $\phi_c$ 
corresponds to a thin barrier.

The surface excitation spectrum, angle-resolved local density of
states, $N(\vp_f,\vR;\epsilon)$
and order parameter, $\Delta(\vp_f,\vR)=
\sqrt{2}\Delta(\vR)(\hat{\vp}_{fx}^2 - \hat{\vp}_{fy}^2)$, are
obtained by solving Eilenberger's transport equations.\cite{eil68}
For high impedance junctions we can neglect
the influence of the tunnel current on the excitation spectrum at
the interface. In order to calculate the surface order parameter
and density of states we solve the transport
equations supplemented by surface boundary conditions.
We consider two models: (i) an atomically 
smooth surface described by continuity of the propagator
at the surface for the incident and specularly
reflected trajectories and,
(ii) a rough or faceted surface modeled as a distribution of microscopic
mirror surfaces that are mis-oriented relative to the average
interface normal.\cite{thu92b}

For specular interfaces the order parameter is strongly suppressed
for a $(110)$ surface because the order 
parameter changes sign along all classical
trajectories. Consequently, an Andreev bound state with
zero energy is formed 
for every trajectory. The formation of bound states comes at
the expense of the continuum states that form the $d_{x^2-y^2}$
pair condensate. Conversely, the specular $(100)$ interface
is not pairbreaking; the order parameter is constant
in magnitude and phase along all incident and specularly
reflected trajectories.
Figure 1a also shows the effect of surface roughness on the
d-wave order parameter for $(110)$ and $(100)$ 
oriented surfaces. For rough surfaces pair-breaking occurs
for all orientations
of the surface normal relative to the crystal axes.
The key feature to note in Figs. 1c-d is the appearance of
a ZBCP with approximately equal spectral weight
for {\it both} $(110)$ and $(100)$ orientations as a
result of surface-induced Andreev scattering by the rough 
surface. Slight differences in the distribution of spectral weight
can be seen as a function of the number of nanofacets.

\begin{figure}[b]
\centerline{
\epsfysize=0.4\textwidth
\rotate[r]{
\epsfbox{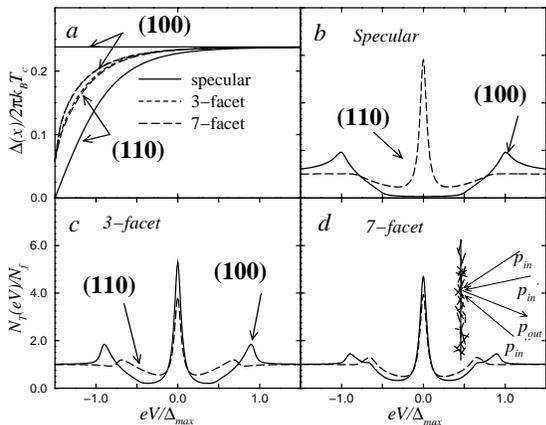}}}
\caption[]{a) The $d_{x^2-y^2}$ order parameter near
$(110)$ and $(100)$ surfaces for specular and rough surfaces.
b) The absence of the ZBCP is a special
feature of a {\it specular} $(100)$ surface.
c) The tunneling conductance for nano-faceted $(110)$
and $(100)$ surfaces with three facets $(0,\pm 45^o$).
d) A similar calculation to c) but with seven facets
$(0,\pm 22.5^o, \pm 45^o,\pm 67.5^o$). The acceptance
cone is $\phi_c=30^o$, the width
parameter is $\gamma=0.05\Delta_0$ and $T=0.3T_c$.
}
\label{Fig1}
\end{figure}

Tunneling experiments performed on oriented YBCO films were
carried out in a magnetic field.\cite{les92,cov96,cov97}
These experiments show a ZBCP which broadens with increasing
field and is split by roughly $3\,meV$ at $H=1\, T$.
The field evolution of the ZBCP in Pb/YBCO films was
initially interpreted in terms of a Zeeman splitting of
resonant magnetic impurities in the tunnel barrier.\cite{app66}
The Zeeman shift of the conductance peak in the Appelbaum theory
is given by $\delta=g\mu_B H$, where $g$ is the g-factor of the paramagnetic
centers. Lesueur, et al.\cite{les92} report a large g-factor
to account for the splitting in Pb/YBCO junctions.
An alternative way of expressing
the Zeeman shift in the Appelbaum model is
$\delta=\Delta_0(H/H_P)$, where $H_P=\Delta_0/g\mu_B$ is the 
Pauli field, i.e. the field scale for pair-breaking by the Zeeman energy.
The data of Lesueur, et al.\cite{les92} implies an 
anomalously low Pauli field ($H_P\sim 10\,T$ vs.
$\Delta_0/2\mu_B=125\,T$) in order to account
for the splitting of the conductance peak in a field.

Here we suggest an explanation of the field evolution
of the ZBCP that does not invoke paramagnetic tunneling
centers, but depends upon
the d-wave interpretation of the ZBCP. The surface bound
states that give rise to the ZBCP couple to the magnetic
field at the interface via the screening current in the superconductor.
The electromagnetic coupling that enters the transport equation is
$-\frac{e}{c}\vv_f\cdot\vA(\vR)$,
where $\vA(\vR)$ is the self-consistently determined vector
potential. For a uniform (or slowly varying) 
supercurrent this coupling
leads to a  Doppler shift in the continuum excitations
given by $\vv_f\cdot\vp_s$, where 
$\vp_s=\left[\frac{\hbar}{2}\grad\vartheta-\frac{e}{c}
\vA(\vR)\right]$ is the condensate momentum,
i.e. excitations co-moving with
the superflow are shifted to higher energy while
counter-moving excitations are shifted to lower energy.
The current also shifts the Andreev bound state spectrum.

Consider the effect of $ab$-plane screening
currents on the tunneling conductance.
The screening current
is parallel to the surface and
proportional to the applied field:
$\vH=H\hat{\vz}$,
$\vp_s=-(e/c)A(x)\hat{\vy}=
(e/c)H\lambda\exp(-x/\lambda)\hat{\vy}$, where
$\lambda$ is the $ab$-plane penetration depth.
First consider a model for the excitation spectrum of a
d-wave superconductor which is not self-consistent, i.e.
neglect pair-breaking of the d-wave order parameter at a 
specular $(110)$ surface. Eilenberger's equation
can then be solved analytically for $\lambda\gg\xi_0$;
the resulting angle-resolved local density of states is given by,
\be
N(\vp_f,x;\epsilon)=\mbox{Im}\left\{
\frac{\tilde{\varepsilon}^R}
{D^R}
-
\frac{|\Delta(\vp_f)|^2}
{\tilde{\varepsilon}^R D^R}
e^{-2D^R\,x/|\vv_f\cdot\hat{\vx}|}
\right\}
\,,
\label{DOS}
\ee
where 
$D^R=\sqrt{|\Delta(\vp_f)|^2-\tilde{\varepsilon}^R(\vp_f,\epsilon)^2}$
and $\tilde{\varepsilon}^R(\vp_f,\epsilon)=
\epsilon + i \gamma + \frac{e}{c}\vv_f\cdot\vA(\vR)$
defines the excitation energy, with impurity broadening approximated 
by a constant width $\gamma$. 
The first term in Eq. \ref{DOS} is the bulk density of states.
The second term gives a bound state contribution near zero energy.
At the interface the total (bulk plus surface) continuum 
contribution to the density of states is given by
$N_{c}=\frac{\sqrt{(\epsilon-\vv_f\cdot\vp_s)^2-|\Delta(\vp_f)|^2}}
{|\epsilon-\vv_f\cdot\vp_s|}\,
\Theta((\epsilon-\vv_f\cdot\vp_s)^2-|\Delta(\vp_f)|^2)$,
which shows the Doppler shift in the continuum edge.
The singularity in the bulk density of states is removed,
and spectral weight from the continuum is shifted to the bound state.
In the limit $\gamma\rightarrow 0$ the bound state contribution becomes
$N_{b}(\vp_f,x;\epsilon)=\pi|\Delta(\vp_f)|\,\delta(\epsilon -
\vv_f\cdot\vp_s)
\,\exp(-2|\Delta(\vp_f)|\,x/|\vv_f\cdot\hat{\vx}|)$,
which is also shifted in energy by the screening current.
The spectral weight decays into the bulk
as the square of the bound state amplitude.
For the trajectory
$\hat{\vv}_f=\cos\phi_{\hat{p}}\hat{\vx}+\sin\phi_{\hat{p}}\hat{\vy}$
the shift in the bound state is
$\epsilon_b=(e/c)v_f H\lambda\sin\phi_{\hat{p}}$. A field
scale is set by a screening current of order the
bulk critical current, $H_0=\frac{c\Delta_0}{ev_f\lambda}/\sin\phi_c$,
where $\phi_c$ determines the maximum shift observable in tunneling.
This field is of order $H_0 \simeq
\frac{\phi_0}{\pi^2\xi_0\lambda}/\sin\phi_c\simeq 1-10\,T$, and
is the field scale observed in the low-field linear region of
the splitting of the ZBCP in Pb/YBCO and Cu/YBCO tunnel junctions.
The shift in the energy of the bound states 
calculated from the London
screening current and Eq. (\ref{DOS}) is
strictly valid for $x\gg\xi_0$,
where the bound states have little spectral weight.
At the interface, where the bound states contribute to the
tunneling conductance, a self-consistent
calculation of the order parameter and current is required.

\begin{figure}[b]
\centerline{
\epsfysize=0.50\textwidth
\rotate[r]{
\epsfbox{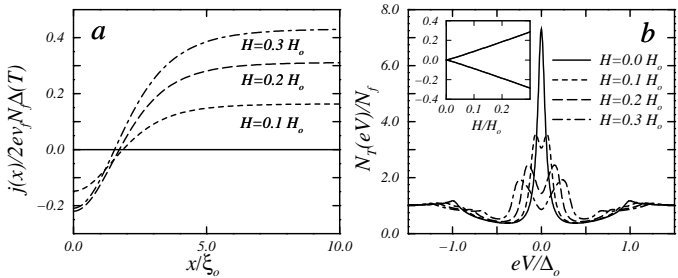}}}
\caption[]{a) The current density
at different applied fields for a pure
$d_{x^2-y^2}$ order parameter
and a specular $(110)$ surface.
The current density heals to the London
screening current, $j\simeq (c/4\pi)H/\lambda$,
for $x > 5\xi_0$. For $x < 5\xi_0$,
the current reverses sign and is
carried by the bound states.
b) The tunneling conductance vs. $H$ for $T=0.3T_c$
and $\phi_c=45^o$.
The splitting of the ZBCP reflects
the shift in the surface bound states by
the screening current at $x=0$.
The inset shows the linear splitting of the ZBCP at low 
fields.
}
\label{Fig2}
\end{figure}

Figure 2a shows the self-consistent result for the
current density for a specular $(110)$ surface
as a function of the applied field $H$ at $T=0.3T_c$.
The current density heals
to the London screening current, $j(x)\simeq (c/4\pi)H/\lambda$, for
$\xi_0\ll x\ll \lambda$, where it is carried
by the continuum states that comprise the
condensate. Near the surface, $x\lesssim\xi_0$,
the current is carried by Andreev
bound states. A similar phenomena is found in
the spectral resolution of the current
density in vortices.\cite{rai96}
The bound-state current scales linearly
with the applied field (in the Meissner region), but 
is {\it counter-flowing} relative to the bulk screening current.
As a result the
bound-states lead to a small increase in the penetration depth,
$\delta\lambda/\lambda_{London}\simeq (\xi_0/\lambda_{London})$.
The bound states also lead to a splitting of the ZBCP in a
magnetic field. The spectrum is calculated
self-consistently with the surface current density.
Figure 2b shows the field evolution of the ZBCP for 
a $d_{x^2-y^2}$ superconductor and a $(110)$ interface.
The splitting is symmetric about zero bias,
with $\delta\simeq \Delta_0(H/H_0)$; 
$H_0 \simeq 6\,T$ is used in the calculations.

Other current sources will also
split the ZBCP. In particular, the
spontaneous currents generated at a surface phase transition
to a pairing state with mixed symmetry, 
e.g. a $d+is$ surface phase, 
are due to a shift in the energies
of the Andreev bound states
in zero field.\cite{mat95,buc95}
The relevance of the sub-dominant order parameter
depends on the microscopic pairing interaction.
Several authors have proposed that superconductivity
in the cuprates is mediated
by the exchange of anti-ferromagnetic (AFM)
spin fluctuations (see Ref.
\onlinecite{sca95} and references therein).
This mechanism predicts d-wave pairing
with $B_{1g}$ symmetry over a wide
range of parameter space of the spin-fluctuation propagator
and bandstructure.
The AFM mechanism also predicts that the
$A_{2g}$ channel $[p_xp_y(p_x^2-p_y^2)]$ is attractive,
but sub-dominant to the $B_{1g}$ channel,
i.e. $0<T_c^{A_{2g}}<T_c^{B_{1g}}$,
and that the pairing interaction in all
other channels is repulsive, including
anisotropic $A_{1g}$ states that have nodes.\cite{buc95}
Solutions to the linearized gap equation with a 
tight-binding bandstructure and
the spin-fluctuation model of Radtke, et al.\cite{rad94}
lead to sizeable values for the $A_{2g}$ coupling strength,
e.g. $T_c^{A_{2g}}/T_c^{B_{1g}}\simeq 0.3-0.4$.\cite{buc95}
Other pairing mechanisms may lead to subdominant 
pairing with different symmetry,
e.g. the electron-phonon interaction may lead to a sub-dominant pairing
interaction in the $A_{1g}$ channel. We consider both possibilities.

Surface pair breaking frees up spectral weight at the Fermi surface
for the formation of pairs in a sub-dominant
pairing channel. For example, a surface order parameter
with $A_{1g}$ symmetry will develop
near a $(110)$ surface as shown 
in the inset of Fig. 3a.\cite{buc95,mat95}
Figure 3b shows the {\it surface} 
phase diagram as a function of the sub-dominant
coupling strength for specular scattering.\cite{buc95}
Below the surface phase transition
temperature, $T_s$,
a surface order parameter develops which
spontaneously breaks ${\cal T}$-symmetry. 
The surface phase is doubly degenerate,
corresponding to two possible directions for
the spontaneous surface current.
For $T_{c2}=0.3T_{c1}$ surface roughness
is found to suppress $T_s$ by roughly $30\%$ for the 
$B_{1g}+iA_{1g}$ transition near a $(110)$ surface.
The magnitude of the spontaneous surface current shown
in Fig. 3a scales with the amplitude of
the subdominant order parameter and can
be a sizeable fraction of the critical current density.
The spontaneous current is carried by the bound states and is
confined to a few coherence lengths of the surface.
The ZBCP splits in zero field as $T$ drops below $T_s$
as shown in Fig. 3c for a $B_{1g}\pm iA_{1g}$ surface phase
and a sub-dominant coupling constant of
$T_{c2}=0.29 T_{c1}$, which gives $T_s=0.25 T_{c1}$,
and a zero-field splitting of $\delta_s \simeq 0.15\Delta_0$.
The zero-field conductance curves calculated 
at points $A$ and $B$ in Fig. 3b are shown in Fig. 3c.
The proximity to the surface phase transition,
leads to a low-field nonlinear evolution of the
splitting of the ZBCP, above and
below $T_s$, as shown in the inset of Fig. 3d.

\begin{figure}[b]
\centerline{
\epsfysize=0.48\textwidth
\rotate[r]{
\epsfbox{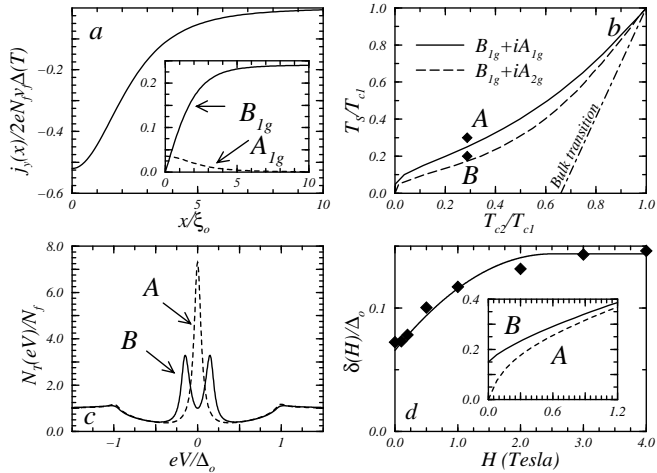}}}
\caption[]{
a) The spontaneous current density of
a $B_{1g}+iA_{1g}$ surface
phase for a $(110)$ surface at $T=0.8T_s$.
The current is scaled in units of the critical current
density. The inset shows the
surface order parameters (in units of $2\pi k_B T_c$).
b) Phase diagram for surface states
with broken ${\cal T}$-symmetry; the solid (dashed) line 
is the instability line for a surface $B_{1g}\pm iA_{1g}$
($B_{1g}\pm iA_{2g}$) state. The bulk transition line is also
shown for both symmetries.
c) Tunneling conductance above and below $T_s$ for
sub-dominant pairing with $A_{1g}$ symmetry. The
zero-field splitting is $\delta_s=0.15\Delta_0$.
d) $\delta(H)$ for $T\ll T_s$.
The diamonds are the data of Ref. 
\onlinecite{cov97}, and the solid line is the theoretical
result with $\delta_s=1.1\,meV$, 
$H_0=16\,T$ and $H_c^*=2.5\,T$.
The inset shows the low field evolution of the
splitting just above and below $T_s$.
}
\label{Fig3}
\end{figure}

Recent experiments on Cu/YBCO tunnel
junctions report a zero-field
splitting of $\delta_s^{expt}=1.1\,meV$,
and a surface phase transition
temperature of $T_s\simeq 7\,K$.\cite{cov97}
These values are in good agreement
with the theoretical predictions for a surface
phase with $B_{1g}+iA_{1g}$ symmetry and
$T_{c2}\simeq 0.15\,T_{c1}$.
This provides evidence for a sub-dominant pairing
channel and surface-induced broken ${\cal T}$-symmetry 
in the high $T_c$ cuprates.
The field evolution of the conductance peaks\cite{les92,cov97}
is also accounted for by the shifts of the
Andreev bound states.
At low temperatures $T\ll T_s$
the low-field nonlinearity is weak, and 
the shift of the ZBCP is given by,
$\delta(H)=\delta_s + v_fp_s\sin\phi_c$;
at low fields $p_s=\frac{e}{c}H\lambda$, and
the splitting increases linearly with $H$.
However, the screening current is nonlinear
at higher fields, and saturates at the
pairbreaking critical current. 
An order of magnitude estimate for the pairbreaking critical
field is given by $p_s\sim\frac{e}{c}H_c\lambda
=\Delta_0/v_f$, i.e.
$H_c\equiv\phi_0/\pi^2\xi_0\lambda\simeq 1\,T$ for
$\lambda=1500\,\AA$ and $\xi_0=15\,\AA$.
The nonlinear correction to
the screening current is dominated by
non-thermal, counter-moving quasiparticles 
near nodal points of the gap,
as described in Ref. \onlinecite{xu94a},
and yields the following
result for the field evolution of the splitting,
$\delta(H)=\delta_s+\Delta_0(H/H_0)
\left[1-\frac{1}{2}(H/H_c^*)\right]$, for $H\le H_c^*$.
This pairbreaking critical field
is determined by the excitations in the vicinity
of the nodes, and is given by
$H_c^*=(\frac{3c}{2e})
d|\Delta(\vartheta)|/d\vartheta|_{\mbox{\small node}} 
/v_f^*$, where
$d|\Delta(\vartheta)|/d\vartheta|_{\mbox{\small node}}$ 
is the angular slope of the $d_{x^2-y^2}$ gap at a node,
and $v_f^*$ is the magnitude of the Fermi velocity at node.
For the standard d-wave model, $|\Delta|\sim |\cos(2\vartheta)|$,
and a cylindrical Fermi surface, $H_c^*=3\,H_c\simeq 3\,T$.
Figure 3d shows a comparison of the data on the
splitting of the ZBCP for Cu/YBCO junctions\cite{cov97}
with the theoretical result for $\delta(H)$ (solid line: 
$H_c^*=2.5\,T$, $H_c=1\,T$ and $\phi_c\simeq 4^o$).

The theory of NIS tunneling applied to d-wave superconductors
accounts for the low-energy features in the $ab$-plane
tunneling conductance in terms of Andreev bound states and 
low-energy continnum excitations near the nodes.
The theory accounts for the  
splitting of the ZBCP in zero-field in terms of
a phase transition to a surface state with
broken ${\cal T}$-symmetry,
and for the field evolution of the conductance
peak in terms of the Doppler shift of the Andreev 
bound states.

We thank L. H. Greene and M. Covington for stimulating
discussions and for sharing their work
with us prior to publication.
We thank J. Kurkij\"arvi for comments on the manuscript.
This work was supported in part by the STC
for Superconductivity through NSF Grant no.~91-20000.
M.F acknowledges partial support from SF{\AA}AF,
{\AA}bo Akademi and Magnus Ehrnrooths Stiftelse.
D.R. and J.A.S. acknowledge support from the
Max-Planck-Gesellschaft and the Alexander von Humboldt-Stiftung.

\end{document}